\documentclass[twocolumn,superscriptaddress,aps,longbibliography,showpacs,preprintnumbers,amsmath,amssymb,floatfix]{revtex4-1}
\usepackage{titlesec}
\usepackage{amsmath}
\usepackage{bm}
\usepackage{graphicx}
\usepackage[ansinew]{inputenc}
\usepackage{array}
\usepackage{color}
\usepackage{amsxtra}
\usepackage{amstext}
\usepackage{amssymb}
\usepackage{latexsym}
\usepackage{gensymb}
\usepackage{dsfont}
\usepackage{braket}
\usepackage[caption=false]{subfig}
\usepackage[makeroom]{cancel}

\usepackage{dcolumn}
\usepackage{bm}
\usepackage{bbm}
\usepackage{braket}
\usepackage{mathrsfs}
\usepackage{amstext}
\usepackage{euscript}
\usepackage{txfonts}

\usepackage[unicode=true,
 bookmarks=false,
 breaklinks=false,pdfborder={0 0 1},backref=false,colorlinks=true,citecolor=blue,urlcolor=blue]
 {hyperref}

\makeatletter
\@ifundefined{textcolor}{}
{%
 \definecolor{BLACK}{gray}{0}
 \definecolor{WHITE}{gray}{1}
 \definecolor{RED}{rgb}{1,0,0}
 \definecolor{GREEN}{rgb}{0,1,0}
 \definecolor{BLUE}{rgb}{0,0,1}
 \definecolor{CYAN}{cmyk}{1,0,0,0}
 \definecolor{MAGENTA}{cmyk}{0,1,0,0}
 \definecolor{YELLOW}{cmyk}{0,0,1,0}
}

\makeatletter
\renewcommand*\env@matrix[1][*\c@MaxMatrixCols c]{%
  \hskip -\arraycolsep
  \let\@ifnextchar\new@ifnextchar
  \array{#1}}
\makeatother

\newcommand{\cref}[1]{Ref.\,\cite{#1}}

\makeatother

\begin{document}

\title{Distribution of Gaussian Entanglement in Linear Optical Systems}
\author{Jiru Liu}
\email{ljr1996@tamu.edu}
\affiliation{Institute for Quantum Science and Engineering (IQSE) and Department of Physics and Astronomy, Texas A\&M University, College Station, TX 77843-4242, USA}

\author{Wenchao Ge}
\affiliation{Institute for Quantum Science and Engineering (IQSE) and Department of Physics and Astronomy, Texas A\&M University, College Station, TX 77843-4242, USA}
\affiliation{Department of Physics, Southern Illinois University, Carbondale, Illinois 62901, USA}

\author{M. Suhail Zubairy}
\affiliation{Institute for Quantum Science and Engineering (IQSE) and Department of Physics and Astronomy, Texas A\&M University, College Station, TX 77843-4242, USA}
\date{\today}

\begin{abstract}
Entanglement is an essential ingredient for building a quantum network that can have many applications. Understanding how entanglement is distributed in a network is a crucial step to move forward. Here we study the conservation and distribution of Gaussian entanglement in a linear network using a new quantifier for bipartite entanglement. We show that the entanglement can be distributed through a beam-splitter in the same way as the transmittance and the reflectance. The requirements on the entangled states and the type of networks to satisfy this relation are presented explicitly. Our results provide a new quantification for quantum entanglement and further insights into the structure of entanglement in a network.
\end{abstract}
\maketitle

\section*{\centering\uppercase\expandafter{\romannumeral1}. Introduction}
Continuous-variables (CV) quantum systems have attracted tremendous interest due to its intriguing properties as well as the remarkable possibilities it can provide in areas such as quantum communication and quantum computing \cite{braunstein2005quantum,lloyd1999quantum,adesso2014continuous,lund2014boson,duan2000inseparability}. As the representative of CV systems, Gaussian states have always been the center of attention for the fact that they are easy to produce \cite{koga2012dissipation} and convenient to operate experimentally \cite{reck1994experimental}.    
\par
Many studies have been reported about witnessing and even quantifying Gaussian entanglement \cite{vedral1997quantifying,vidal2002computable,simon2000peres,horodecki2001separability,peres1996separability,giedke2001separability,nakano2013negativity,plenio2005logarithmic}. Positive partial transpose (PPT), which gives a direct criterion of separability \cite{simon2000peres,horodecki2001separability,peres1996separability,giedke2001separability}, is widely used to determine the entanglement of bipartite Gaussian systems. Negativity $\mathcal{N}$ and logarithmic negativity $E_\mathcal{N}$ \cite{nakano2013negativity,plenio2005logarithmic} are constructed to quantify bipartite entanglement based on PPT criterion. In a multipartite Gaussian system, structures of how quantum correlations are shared among many parties, such as monogamy property \cite{Terhal04}, have been studied \cite{koashi2004monogamy,Adesso_2006, hiroshima2007monogamy,de2014monogamy}.
In particular, monogamy inequalities have been proved using entanglement quantifications related to negativity and logarithmic negativity \cite{Adesso_2006, hiroshima2007monogamy}. For example, for a generic tripartite system $\rho_{ABC}$, the following inequality is satisfied:
\begin{equation} \label{monogamy_intro}
\mathcal{E}(A,BC)\geqslant \mathcal{E}(A,B)+\mathcal{E}(A,C),
\end{equation}
where $\mathcal{E}(A,BC)$ represents the \emph{one-mode-vs-two-mode} entanglement
among the bipartition $A:BC$ and $\mathcal{E}(A,B)$ ($\mathcal{E}(A,C)$) is the bipartite entanglement of the reduced system after tracing the party $C$ ($B$). Within the monogamy inequality, the summation of entanglement is proposed.
However, it remains an open question which measure can be the 'proper candidate for approaching the task of quantifying entanglement sharing in CV system' \cite{Adesso_2006}. The difficulties lie in the restriction of each measure. For example, Eq. (\ref{monogamy_intro}) can be violated with the logarithmic negativity measure (examples in Sec.\uppercase\expandafter{\romannumeral3} C), which is used widely as the entanglement measure in CV systems \cite{vidal2002computable,plenio2005logarithmic}. 
\par
Gaussian entanglement can be generated in linear-optical networks consisting beam-splitters (BS) by transforming nonclassical non-entangled Gaussian states to entangled ones \cite{ge2018distributed,tahira2009entanglement,tahira2011gaussian,paris1999entanglement,marian2001inseparability}. BS linear networks have been studied in demonstrated distributed quantum metrology tasks \cite{ge2018distributed}, quantum  computing  supremacy \cite{zhong2020quantum} and quantum communication \cite{salih2013protocol}. Recently, different conservation relations of single-mode nonclassicality and two-mode entanglement between the input states and the output states have been reported \cite{ge2015conservation,vogel2014unified}. However, such a conservation or distribution of entanglement in optical systems has not been studied, for example, in linear-optical systems with two or more BSs. 
\par
In this paper, we study the distribution pattern of Gaussian entanglement in a linear network consisting of multiple BSs with single-mode Gaussian states being the inputs. First, we investigate the difference of single-mode nonclassicalities before and after going through a BS, termed as 'residual nonclassicality', in a linear network. It has been shown that the residual nonclassicality can quantify two-mode entanglement and form a conservation relation of nonclassicality and entanglement in a single BS system \cite{ge2015conservation,tasgin2020nonclassicality}. Thus, it is an interesting question whether the residual nonclassicality could be extended to a linear optical system with multiple modes. However, our results show that it does not reveal how entanglement is distributed when the system becomes more complex. Therefore, we propose a new bipartite entanglement quantifier $\xi$, which is defined via logarithmic negativity, and study entanglement distribution using this new quantifier. Our results show that bipartite entanglement distributed through a BS can follow the same relation as how light is distributed at the BS. Based on this relation, we obtain a monogamy equality of a tripartite Gaussian state in a network of two BSs, where the tripartite state is generated from a two-mode entangled state mixed with a single-mode quantum state at a BS. We identify the conditions of the input states in order for the equality to hold. The distribution of entanglement is further extended to more complex networks using the new quantifier $\xi$.
\par
The paper is organized as follows. In Sec.\uppercase\expandafter{\romannumeral2}, we introduce the properties of Gaussian states in a BS optical system as well as three definitions of entanglement quantification. In Sec.\uppercase\expandafter{\romannumeral3}, we derive the conservation relation and distribution pattern of entanglement on the base of the quantifier $\xi$ and connect it to the monogamy of quantum entanglement within various BS systems. Numerical examples are shown in Sec.\uppercase\expandafter{\romannumeral3} C. A summary and future insight are discussed in Sec.\uppercase\expandafter{\romannumeral4}. Detailed derivations are provided in the Appendices. 
\section*{\centering\uppercase\expandafter{\romannumeral2}. Gaussian state entanglement preliminaries}
In this section, we first introduce some basics of single-mode Gaussian states as well as nonclassicality in terms of the characteristic function and the covariance matrix. Then we discuss entanglement quantification of two-mode Gaussian states using logarithmic negativity, residual nonclassicality, and a new entanglement quantifier. 

\subsection*{A. Gaussian state and nonclassicality}
A quantum state $\rho$ can be completely described by the characteristic function \cite{weedbrook2012gaussian}
\begin{equation}
    \chi(\boldsymbol{x})=\text{Tr}[\rho D(\boldsymbol{x})],
\end{equation}
where 
$D(\boldsymbol{x})=\exp[-i\sum_{k=1}^n(q_k\hat{X}_k+p_k\hat{P}_k)]$, $\boldsymbol{x}^T=(q_1,p_1...q_n,p_n)$ and $\hat{X}_k,\hat{P}_k$ are the space-momentum operators. 

A Gaussian state \cite{giedke2001separability,weedbrook2012gaussian} is defined such that its characteristic function is Gaussian (Appendix A).  For a single-mode Gaussian state, its characteristic function is given by
\begin{equation}
\chi(\alpha^*,\alpha)=\text{exp}(-\frac12\boldsymbol{x'}^\dag V \boldsymbol{x'}),
\end{equation}
with
\begin{equation*}
\boldsymbol{x'}^\dag=(\alpha^*,\alpha),\ \text{and}\ V=\left[
  \begin{matrix}
   a & b \\
   b^* & a
  \end{matrix}\right].  
\end{equation*}
In the above equations, $V$ is the covariance matrix with $a^2-|b|^2\ge1/4$ from the uncertainty principle. Instead of quardrature field variables $(q,p)$, we use bosonic field expression $(\alpha^*,\alpha)$ here (see Appendix A for transformation relations).
\par
The nonclassicality of a quantum state $\rho$ can be quantified by the nonclassical depth $\tau$, which, for a Gaussian state, is related to \cite{wolf2003entangling,tasgin2020quantifications} the minimum eigenvalue $\lambda=a-|b|$ of its covariance matrix $V$ as $\tau=\text{max}\{0,1/2-\lambda\}$.
A quantum state is nonclassical if $\tau>0$ or $\lambda<1/2$. Alternatively, we can consider the quantity 
\begin{equation} \label{nonclassicality}
  N=-\log_2(2\lambda)  
\end{equation}
as the nonclassicality of a single-mode Gaussian state \cite{ge2015conservation}. In contrast to $\lambda$, $\Lambda=a+|b|$ is the maximum eigenvalue which will be used later.

\subsection*{B. Entanglement quantification of two-mode Gaussian states}
We consider a lossless BS with two single-mode Gaussian fields being the inputs as shown in Fig. 1. We denote $\cos^2\theta$ as the transmittance and $\varphi$ as the phase shift of the BS. For two separable single-mode Gaussian states, their combined characteristic function is given by
\begin{equation}
\chi_{\text{in}}(\alpha_1^*,\alpha_1,\alpha_2^*,\alpha_2)=\text{exp}(-\frac12\boldsymbol{x'}^\dag V_{\text{in}} \boldsymbol{x'}),
\end{equation}
where $\boldsymbol{x'}^\dag=(\alpha^*_1,\alpha_1,\alpha^*_2,\alpha_2)$, and $$V_\text{in}=
  \left[
  \begin{matrix}
   A & 0 \\
   0 & B
  \end{matrix} 
  \right],
  \ A=\left[
  \begin{matrix}
   a & b \\
   b^* & a
  \end{matrix} 
  \right],
  \ B=\left[
  \begin{matrix}
   c & d \\
   d^* & c
  \end{matrix} 
  \right].
$$
The covariance matrix of the output field, $V_\text{out}$, is derived by a unitary transformation of $V_\text{in}$ as
\begin{equation} \label{Vout}
    V_\text{out}=U^\dag(\theta,\varphi)V_\text{in}U(\theta,\varphi).
\end{equation}
Here we give the exact form of $U(\theta,\varphi)$,
\begin{equation} \label{U}
   \left[
  \begin{matrix}
   \cos\theta & 0 & \sin\theta e^{i\varphi} &0 \\
   0 & \cos\theta & 0 & \sin\theta e^{-i\varphi} \\
   -\sin\theta e^{-i\varphi} & 0 & \cos\theta & 0 \\
   0 & -\sin\theta e^{i\varphi} & 0 & \cos\theta
  \end{matrix} 
  \right].
\end{equation}
\begin{figure}
\begin{center}
\includegraphics[width=5cm,height=5cm]{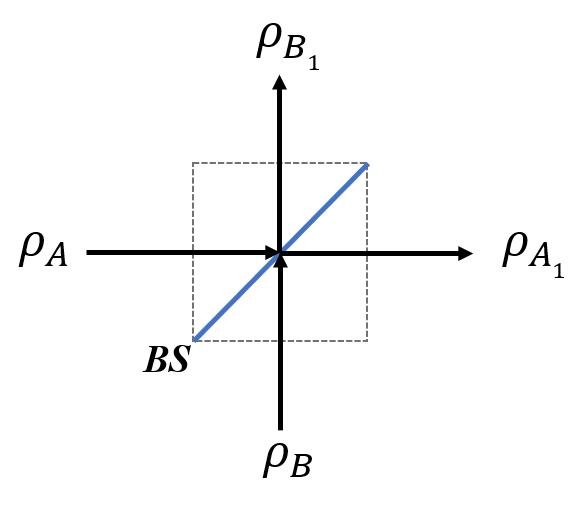}
\caption{A nonclassical Gaussian state $\rho_A$ is mixed with another state $\rho_B$ at a BS, generating the output state $\rho_{A_1B_1}$. The output modes $\rho_{A_1}$ and $\rho_{B_1}$can be obtained by tracing out the mode $B$ and the mode $A$, respectively, on $\rho_{A_1B_1}$.}
\end{center}
\end{figure}
\\
\subsubsection*{B1. Logarithmic negativity $E_\mathcal{N}$}
The entanglement of the two-mode system $\rho_{\text{out}}$ after the BS can be evaluated using the logarithmic negativity $E_\mathcal{N}=\log_2||\rho_{\text{out}}^{T_A}||_1$, where $||R||_1$ denotes the trace norm $\text{Tr}\sqrt{R^{\dagger}R}$ and $\rho^{T_A}$ is the partial transpose of a state.
The condition for two-mode Gaussian state to be entangled is $E_\mathcal{N}>0$.
\par
For Gaussian states, entanglement can be determined totally by its covariance matrix (CM) instead of the density matrix \cite{werner2001bound}. Logarithmic negativity  can be calculated from CM directly, which is much more convenient to manipulate. The logarithmic negativity for the output Gaussian state $\rho_{A_1B_1}$ in Fig. 1 is given by \cite{vidal2002computable}
\begin{equation}\label{log_neg}
    E_\mathcal{N}=\max\bigg\{0,-\frac12\log_2(S-\sqrt{S^2-16\ \text{Det}[V_\text{out}}]\bigg\},
\end{equation}
where $S=2\ \text{Det}[A_1]+2\ \text{Det}[B_1]-4\ \text{Det}[(AB)_1]$, and $A_1,B_1,(AB)_1,(AB)_1^\dag$ are two-dimensional matrices coming from the output covariance matrix $V_\text{out}=\left[
    \begin{matrix}
    A_1 & (AB)_1 \\
    (AB)_1^\dag & B_1
    \end{matrix}
    \right]$ (see details in Appendix C).

\subsubsection*{B2. Residual nonclassicality $S_\mathcal{N}$}
Alternatively, the degree of entanglement of the two-mode Gaussian state can be quantified via the difference between the nonclassicalities before and after the BS \cite{ge2015conservation}. This quantity is denoted as residual nonclassicality $S_\mathcal{N} \equiv N_{in}-N_{out}$, where the subscripts $in$ and $out$ denote the total nonclassicality of the input modes and the output modes, respectively. With the definition in Eq. (\ref{nonclassicality}), we obtain 
\begin{equation} \label{S_N}
S_\mathcal{N}=N_A+N_B-N_{A_1}-N_{B_1}\\
= \log_2\frac{\lambda_{A_1}\lambda_{B_1}}{\lambda_A\lambda_B},    
\end{equation}
where $\lambda_{A_1}=\cos^2\theta\cdot\lambda_A + \sin^2\theta\cdot\lambda_B$ and $\lambda_{B_1}=\sin^2\theta\cdot\lambda_A + \cos^2\theta\cdot\lambda_B$ (Appendix C).
Since BS does not create extra nonclassicality as a linear optical device, $S_\mathcal{N}$ can be related to the degree of entanglement. In fact, it is shown that \cite{ge2015conservation} 
\begin{equation}
    S_\mathcal{N}>0 \quad \Longleftrightarrow \quad E_\mathcal{N}>0
\end{equation}
for certain input Gaussian states (see Appendix C for detailed proof) with certain constraints.
\par
Those constraints, which are explained in details in Appendix C, include
\begin{equation} \label{constraints_1}
\begin{aligned}
&(a)\ \text{both of $\rho_A$ or $\rho_B$ are pure states},\qquad\qquad\qquad\\
&(b)\ \varphi=\frac12[\text{arg}(b)-\text{arg}(d)],\qquad\qquad\qquad\\
\end{aligned}\\
\end{equation}

\subsubsection*{B3. Entanglement quantifier $\xi$}
From Eq. (\ref{log_neg}), it can be derived that $E_\mathcal{N}>0$ is equivalent to
\begin{equation} \label{xi_expression}
    \xi\equiv S-\frac12-8\ \text{Det}[V_\text{out}]>0,
\end{equation} 
where the expression for $S$ can be calculated as
\begin{equation} \label{S}
    S=\frac1{2||\rho_\text{out}^{T_A}||^2_1}+8\text{Det}[V_\text{out}]\cdot||\rho_\text{out}^{T_A}||^2_1.
\end{equation}
The purity of two-mode Gaussian state follows $||\rho_\text{out}^2||_1=\frac14|V_\text{out}|^{-\frac12}$. Together with Eq. (\ref{xi_expression},\ref{S}) we obtain
\begin{equation} \label{eq:xi-def}
    \xi=\frac12\left(1-\frac1{||\rho_\text{out}^{T_A}||^2_1}\right)\left(\frac{||\rho_\text{out}^{T_A}||^2_1}{||\rho_\text{out}^2||^2_1}-1\right).
\end{equation}
The above expression shows that $\xi>0$ is equivalent to the PPT criterion in terms of determining whether entanglement exists. This is a sufficient and necessary condition for an arbitrary two-mode Gaussian state to be entangled \cite{duan2000inseparability, simon2000peres}. In the following, we show that this new quantifier can lead to an entanglement conservation and a distribution relation in a linear network.

\begin{figure}[h]
\begin{center}
\includegraphics[width=5.5cm,height=5.5cm]{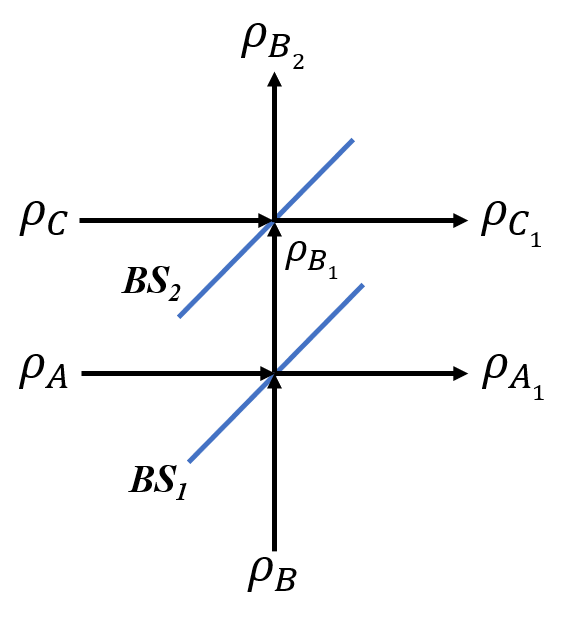}
\caption{
Based on Fig. 1, the Gaussian state $\rho_{B_1}$ mixes with another Gaussian state $\rho_C$ via $BS_2$, generating  $\rho_{B_2C_1}$. With trace operation $\text{Tr}_C$ and $\text{Tr}_B$ on $\rho_{B_2C_1}$, we can obtain $\rho_{B_2}$ and $\rho_{C_1}$, respectively. The whole system of the mode A, B and C after the two BSs is represented by $\rho_\text{out}=\rho_{A_1B_2C_1}$, in which case $\rho_{B_2C_1}=\text{Tr}_A[\rho_\text{out}]$.}
\end{center}
\end{figure}

\section*{\centering\uppercase\expandafter{\romannumeral3}. Distribution of entanglement in linear optical systems}

\subsection*{A. Two BS system}
In this section, we investigate how bipartite entanglement is distributed in a linear system with two beam-splitters as shown in Fig. 2. Two single-mode Gaussian states are mixed at the first BS to generate a bipartite entangled state $\rho_{A_1B_1}$. After that, one of the output mode $\rho_{B_1}$ is mixed with the third single-mode input Gaussian state $\rho_C$ at the second BS. The final output state is a tripartite Gaussian state $\rho_\text{out}$ (or $\rho_{A_1B_2C_1}$). In general, there are both bipartite entanglement and tripartite entanglement in the system after the two BSs. 

\subsubsection*{A1. Distribution of entanglement via residual nonclassicality}
An interesting question is to understand how nonclassicality is shared within the system. As described by the caption in Fig. 2, we can apply Eq. (\ref{S_N}) at both beam-splitters individually to obtain
\begin{equation} \label{S_N A1B1}
    N_A+N_B=N_{A_1}+N_{B_1}+S_\mathcal{N}(A_1,B_1),
\end{equation}
and
\begin{equation} \label{S_N B2C1}
    N_{B_1}+N_C=N_{B_2}+N_{C_1}+S_\mathcal{N}(B_2,C_1).
\end{equation}
The summation of Eq. (\ref{S_N A1B1}) and Eq. (\ref{S_N B2C1}) leads to
\begin{equation}
    N_A+N_B+N_C=N_{A_1}+N_{B_2}+N_{C_1}+S_\mathcal{N},
\end{equation}
with
\begin{equation*}    
    S_\mathcal{N}=S_\mathcal{N}(A_1,B_1)+S_\mathcal{N}(B_2,C_1),
\end{equation*}
where $S_\mathcal{N}$ quantifies the difference of nonclassicality before and after the two BSs, which is given by
\begin{equation}
\begin{aligned}
    S_\mathcal{N}&=\log_2\frac{\lambda_{A_1}\lambda_{B_1}}{\lambda_A\lambda_B}+\log_2\frac{\lambda_{B_2}\lambda_{C_1}}{\lambda_{B_1}\lambda_C} \\
    &=\log_2\frac{\lambda_{A_1}\lambda_{B_2}\lambda_{C_1}}{\lambda_A\lambda_B\lambda_C},
\end{aligned}
\end{equation}
where the two terms in the first line agree with $S_\mathcal{N}(A_1,B_1)$ and $S_\mathcal{N}(B_2,C_1)$, respectively. $S_\mathcal{N}(B_2,C_1)$ stands for the entanglement of $\rho_{B_2:C_1}$  while $S_\mathcal{N}(A_1,B_1)$ stands for the entanglement of $\rho_{A_1:B_1}$, or the entanglement of $\rho_{A_1:B_2C_1}$. In order to explore how the bipartite entanglement of $\rho_{A_1:B_2C_1}$ is distributed at $BS_2$, we need to quantify these two contributions from the entanglement of $\rho_{A_1:B_2},\ \rho_{A_1:C_1}$. However, quantifying the entanglement of $\rho_{A_1:B_2}$, $\rho_{A_1:C_1}$ in terms of the residual nonclassicality is not straightforward since both the two bipartite systems are not directly generated from two separable modes. Therefore, we seek for an alternative solution using another quantifier for entanglement, which is $\xi$ as we introduced in Eq. \eqref{eq:xi-def}. 

\subsubsection*{A2. Distribution relation of entanglement via $\xi$}
Using the entanglement quantifier $\xi$, we denote the  bipartite  entanglement of the states $\rho_{A_1:B_1},\rho_{A_1:B_2}$ and $\rho_{A_1:C_1}$ as $\xi(A_1,B_1), \xi(A_1,B_2)$ and $\xi(A_1,C_1)$, respectively.  
\par
We derive a mathematical result based on the expression of $\xi(A_1,B_1), \xi(A_1,B_2)$ and $\xi(A_1,C_1)$. The three quantities satisfy the following relation (see Appendix D for details, the calculation tricks are introduced in Appendix B) 
\begin{equation} \label{distribution_1}
\begin{aligned}
\xi(A_1,B_2) =& \cos^2{\theta_2}\cdot \xi(A_1,B_1),\\
\xi(A_1,C_1) =& \sin^2{\theta_2}\cdot \xi(A_1,B_1),\\
\end{aligned}
\end{equation}
where $\cos^2{\theta_2}$ is the transmittance of $BS_2$. With the new entanglement quantifier $\xi$, it exactly shows that entanglement could be distributed by a BS in the way how transmittance and reflectance are distributed, which are $\cos^2\theta$ and $\sin^2\theta$, respectively. 

We discover the type of input states and the requirements of a linear network in order for the above distribution relation to hold. These constraints are summarized as follows:
\begin{equation} \label{constraints_2}
\begin{aligned}
(a)&\ \text{$\rho_A,\rho_B,\rho_C$ should all be pure Gaussian states}. \\
(b)&\ \text{The phase shift of $BS_2$},\ \varphi_2=\frac12[\arg(d)-\arg(f)], \\ &\text{where}\ f\ \text{comes from CM}\ V_C=\left[\begin{matrix}
g & f\\
f^* &f
\end{matrix}\right] .\\
\end{aligned}
\end{equation}

\subsection*{B. Monogamy of quantum entanglement}
Considering the tripartite Gaussian state $\rho_{A_1B_2C_1}$, we reveal further a connection between the distribution relation and the monogamy of entanglement. 
\par
For the configuration in Fig. 2, the entanglement of $\rho_{A_1:B_2C_1}$, denoted as $\mathcal{E}(A_1,B_2C_1)$, is related to the entanglement of $\rho_{A_1:B_1}$, denoted as $\mathcal{E}(A_1,B_1)$. We prove that (Appendix E)
\begin{equation} \label{tracenorm_eq}
\begin{aligned}
    ||\rho_{A_1:B_2C_1}^{T_A}||_1&=||\rho_{A_1:B_1}^{T_A}||_1, \\
    ||\rho_{A_1:B_2C_1}^{2}||_1&=||\rho_{A_1:B_1}^{2}||_1,
\end{aligned}
\end{equation}
based on which, both the entanglement quantifiers $\xi$ defined in Eq. (\ref{eq:xi-def}) and logarithmic negativity $E_\mathcal{N}$ defined in Eq. (\ref{log_neg}) provide that
\begin{equation} \label{two_en_equ}
    \mathcal{E}(A_1,B_2C_1)=\mathcal{E}(A_1,B_1),
\end{equation}
\par
When entanglement $\mathcal{E}$ is quantified with $\xi$ measure, together with the results in Eq. (\ref{distribution_1}) and Eq. (\ref{two_en_equ}), it is given that
\begin{equation}
    \mathcal{E}(A_1,B_1)=\mathcal{E}(A_1,B_2)+\mathcal{E}(A_1,C_1),
\end{equation}
which means
\begin{equation} \label{monogamy_equlity}
    \mathcal{E}(A_1,B_2C_1)=\mathcal{E}(A_1,B_2)+\mathcal{E}(A_1,C_1).
\end{equation}
On the other hand, monogamy of quantum entanglement is expressed as \cite{koashi2004monogamy,de2014monogamy} 
\begin{equation} \label{monogamy}
\mathcal{E}(A,BC)\geqslant \mathcal{E}(A,B)+\mathcal{E}(A,C).
\end{equation}
It implies that the conservation relation in Eq. (\ref{monogamy_equlity}) is actually a special case when monogamy inequality becomes an equality, which means in the most case, there is no such a conservation relation. Thus in Eq. (\ref{constraints_2}), conditions and constraints must be employed to make sure such an equality holds. Besides of the trivial constraint about the phase shift, the pure state condition is the only thing left, which we believe to be of enough generality to be accepted. For the case where one of the constraints is not satisfied, to be specific, $\rho_C$ is not a pure state, the conservation equality is destroyed and becomes (according to the proof procedures in Appendix D)
\begin{equation}
    \mathcal{E}(A_1,B_1)>\mathcal{E}(A_1,B_2)+\mathcal{E}(A_1,C_1),
\end{equation}
which agrees with the monogamy inequality in Eq. (\ref{monogamy}). 
\\
\\
\begin{figure*}
\begin{center}
\includegraphics[width=14cm,height=10cm]{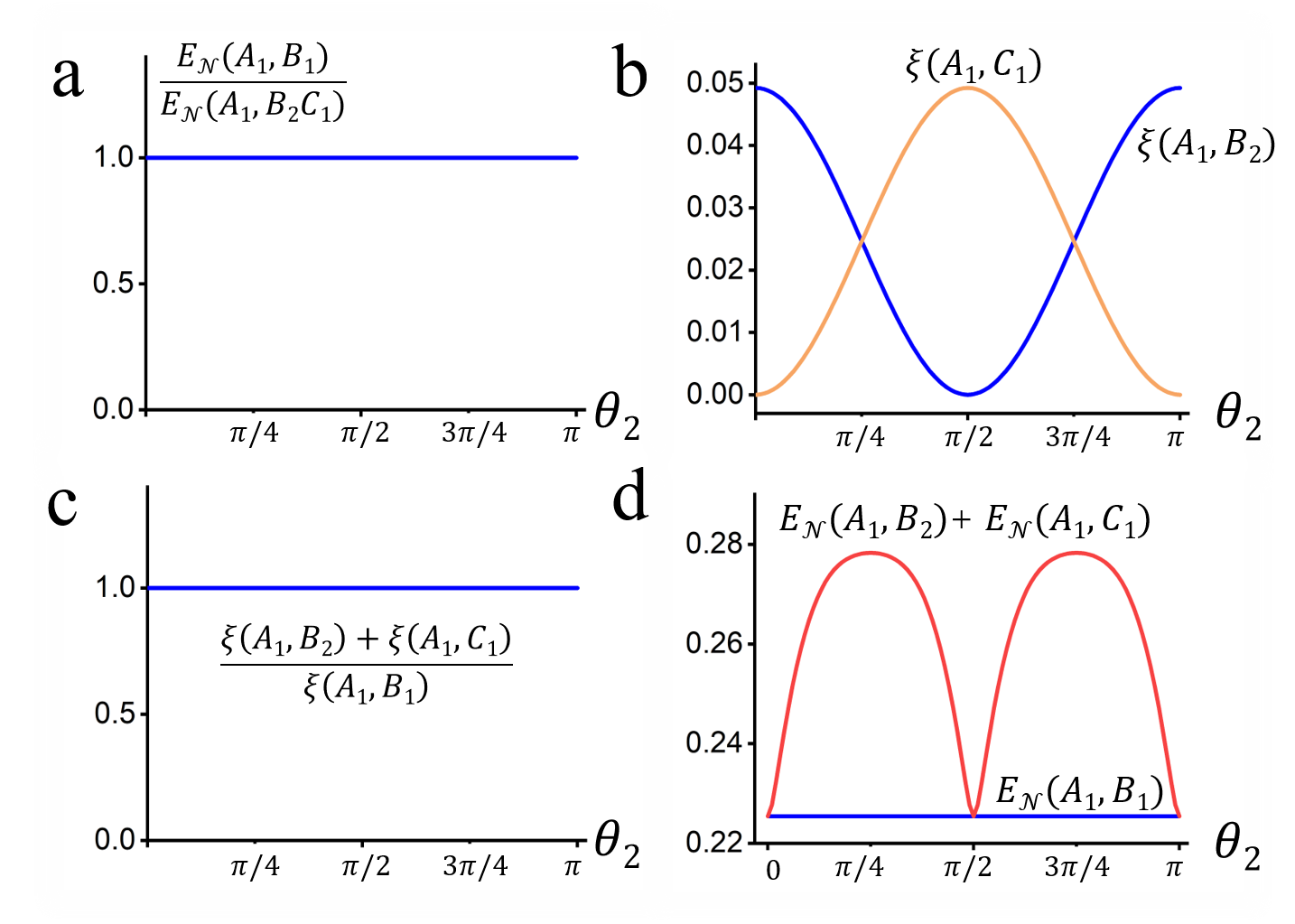}
\caption{\textbf{a.} The ratio of $E_\mathcal{N}(A_1,B_1)/E_\mathcal{N}(A_1,B_2C_1)$ versus $\theta_2$.  \textbf{b.} $\xi(A_1,B_2)$ and $\xi(A_1,C_1)$ versus $\theta_2$. $\xi(A_1,B_2)$ acts in a $\cos^2\theta_2$ relation  while $\xi(A_1,C_1)$ presents a $\sin^2\theta_2$ dependence. \textbf{c.} The ratio of $[\xi(A_1,B_2)+\xi(A_1,C_1)]/\xi(A_1,B_1)$ versus $\theta_2$. \textbf{d.} $E_\mathcal{N}(A_1,B_1)$ and the summation of $E_\mathcal{N}(A_1,B_2)+E_\mathcal{N}(A_1,C_1)$ versus $\theta_2$.}
\end{center}
\end{figure*}
\subsection*{C. Comparison between $E_\mathcal{N}$ and $\xi$}
The most important point of the quantifier $\xi$ lies in the fact it provides a precise $\cos^2\theta$-$\sin^2\theta$ distribution pattern. A conservation relation is also naturally satisfied. However, these doesn't happen when using the measure of logarithmic negativity $E_\mathcal{N}$.  
\par

In Fig. 3, we show the value of $E_\mathcal{N}$ and $\xi$ versus $\theta_2$ ($BS_2$). The results describe the configuration in Fig. 2, where all the parameters are given proper numerical values: For the CM of three pure Gaussian states, $a=\sqrt{0.34},b=0.3,c=\sqrt{0.5},d=0.5,g=\sqrt{0.74},f=0.7$. For $BS_1$, $\theta_1=\pi/4,\phi_1=0$. For $BS_2$, $\phi_2=1,\theta_2$ varies from $0$ to $\pi$. In Fig. 3a, $E_\mathcal{N}(A_1,B_2C_1)$ is the \emph{one-mode-vs-two-mode} entanglement calculated by the minimum symplectic eigenvalue \cite{li2018magnon} of the three-mode Gaussian state CM, which is the actually the same thing as Eq. (\ref{log_neg}). The plot shows clearly that $E_\mathcal{N}(A_1,B_1)=E_\mathcal{N}(A_1,B_2C_2)$, which agrees with Eq. (\ref{tracenorm_eq}). In Fig. 3b, $\xi(A_1,B_2)$ and $\xi(A_1,C_1)$, which are calculated through Eq. (\ref{xi_expression}), vary with $\theta_2$, present a $\cos^2\theta_2,\sin^2\theta_2$ dependence to $\theta_2$, respectively. This entanglement distribution pattern is exactly what we get in Eq. (\ref{distribution_1}). Fig 3c shows that $\xi(A_1,B_2)+\xi(A_1,C_1)=\xi(A_1,B_1)$, together with Eq. (\ref{two_en_equ}), finally proves the monogamy equality $\xi(A_1,B_2C_1)=\xi(A_1,B_2)+\xi(A_1,C_1)$. In Fig 3d, logarithmic negativity is used as the entanglement measure. It is clear that it violates the monogamy inequality in Eq. (\ref{monogamy}) and fails in revealing the distribution pattern of entanglement. 
\par

\subsection*{D. Complex BS networks}
The distribution relation of entanglement in the tripartite system can be extended to muti-partitle systems in some more complex linear optical networks.
\begin{figure}[h]
\begin{center}
\includegraphics[width=5.5cm,height=6.5cm]{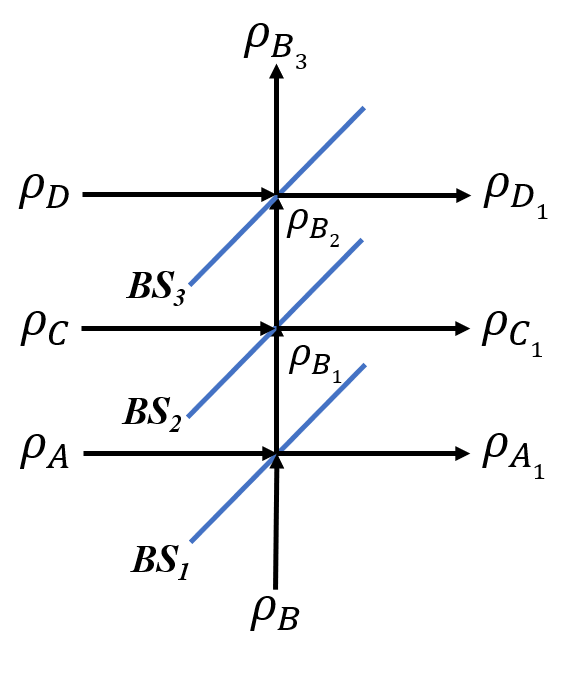}
\caption{
Based on Fig. 2, another beam-splitter $BS_3$ and another input state $\rho_D$ are added to the network. The CM of $\rho_D$ is given by $V_D=\left[\begin{matrix}
h & j\\
j^* &h
\end{matrix}\right]$.}
\end{center}
\end{figure}
\par
We consider a four-mode Gaussian state generated by a three-BS network (Fig. 4). As shown by the figure, three BSs are linearly arranged. Detailed calculation (Appendix F) indicates that 
\begin{equation} \label{distribution_2}
\begin{aligned}
\xi(A_1,B_3) &= \cos^2{\theta_3}\cdot \xi(A_1,B_3D_1), \\
\xi(A_1,D_1) &= \sin^2{\theta_3}\cdot \xi(A_1,B_3D_1),
\end{aligned}
\end{equation}
where $\xi(A_1,B_3D_1)=\xi(A_1,B_2)$. The distribution relation still works here and it is easy to write down the exact value of entanglement between $A_1$ and any other states. For $\xi(A_1,D_1)$, combining Eq. (\ref{distribution_1}) and Eq. (\ref{distribution_2}), it is given that
\begin{equation}
\xi(A_1,D_1) = \sin^2\theta_3\cos^2\theta_2\cdot\xi(A_1,B_1).
\end{equation}
The summation of all the distribution relation equations provides a conservation equality of entanglement in the four-mode states as well, which is given by
\begin{equation} \label{conservation_twobs}
\xi(A_1,B_3)+\xi(A_1,D_1)+\xi(A_1,C_1)=\xi(A_1,B_3D_1C_1),
\end{equation}
where $\xi(A_1,B_3C_1D_1)=\xi(A_1,B_1)$.
\par
Constraints for the distribution relation to hold in Fig. 4 include:
\begin{equation}
\begin{aligned}
    &(b)\ \text{$\rho_A,\rho_B,\rho_C,\rho_D$ should all be pure Gaussian states}.\\
    &(b)\ \text{The phase shift of $BS_3$},\ \varphi_3=\frac12[\arg(f)-\arg(j)].\\
    &(c)\ \text{The covariance matrix of $\rho_C$ equals to the one of $\rho_D$.}
\end{aligned} 
\end{equation}
As demanded by the last constraint, $V_C=V_D$, both input states C and D can be chosen to be vacuum for simplicity during experimental implementation, which means
\begin{equation}
    V_C=V_D=\left[
        \begin{matrix}
            0.5 &0\\
            0&0.5
\end{matrix}\right].
\end{equation} 
\begin{figure}[h]
\begin{center}
\includegraphics[width=5.5cm,height=7cm]{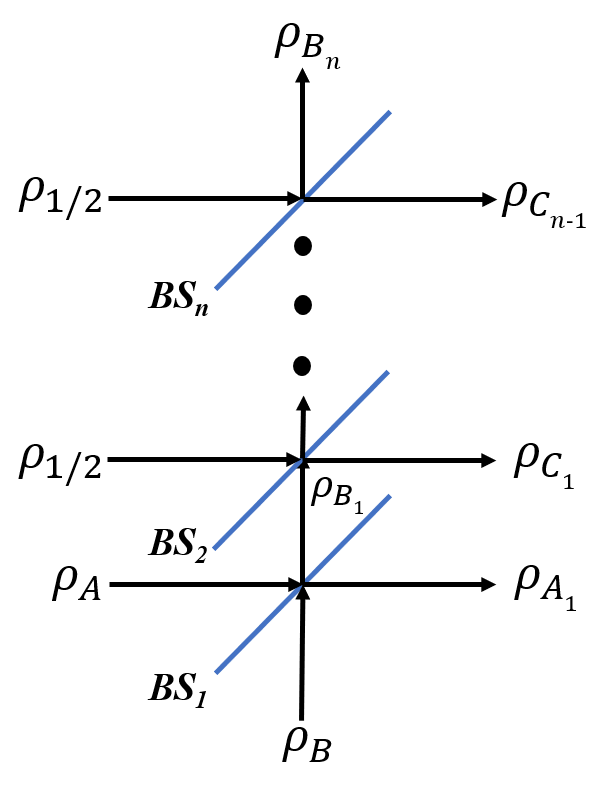}
\caption{
Linearly arranged BS network with $n$ BSs, where $\rho_{1/2}$ stands for vacuum or coherent states and its CM $V_{1/2}=\frac12\mathbf{I_2}$.}
\end{center}
\end{figure}
\par
We then generalize the distribution relation for a $(n+1)$-mode Gaussian state created under the configuration in Fig. 5. All the inputs state except for $\rho_A,\rho_B$ are chosen as vacuum or coherent states. In this case, the distribution relation provides 
\begin{equation}
\begin{aligned}
    \xi(A_1,B_n)&=\prod_{i=2}^n \cos^2\theta_i\cdot \xi(A_1,B_nC_1C_2...C_{n-1}), \\
    \xi(A_1,C_k)&=\sin^2\theta_{k+1}\prod_{i=2}^{k} \cos^2\theta_i\cdot \xi(A_1,B_nC_1C_2...C_{n-1}),
\end{aligned}
\end{equation}
where $\xi(A_1,B_nC_1C_2...C_{n-1})=\xi(A_1,B_1)$ and $n>1,\ k>1$. Applying the same method as we derive Eq. (\ref{conservation_twobs}), the conservation relation is given by
\begin{equation}
\xi(A_1,B_n)+\sum_{i=1}^{n-1}\xi(A_1,C_i)=\xi(A_1,B_nC_1C_2...C_{n-1}).
\end{equation}

\section*{\centering\uppercase\expandafter{\romannumeral4}. Discussion and conclusion}
In this paper, we discussed the distribution of Gaussian entanglement in linear-optical networks. A new quantifier of bipartite entanglement is introduced, based on which we show that the monogamy inequality and even an monogamy equality are satisfied. For pure input states, entanglement can be a conserved quantity. Moreover, it obeys the distribution pattern in the same way how transmittance and reflectance are distributed when going through a BS. Such property makes it possible to calculate and even control the entanglement of any two subsystems, which provides us a deeper understanding about monogamy of quantum entanglement and how entanglement is distributed in a quantum network. 
\par
The new entanglement quantifier $\xi$, which is closely related to logarithmic negativity and residual nonclassicality, proves better properties than both measure in monogamy inequality verification and entanglement distribution. It can contribute to the research on \emph{residual entanglement} \cite{Adesso_2006,coffman2000distributed,li2018magnon} $\mathcal{R}$ given by \cite{Adesso_2006}
\begin{equation}
    \mathcal{R}(A:BC)=\mathcal{C}(A:BC)-\mathcal{C}(A:B)-\mathcal{C}(A:C),
\end{equation}
where $\mathcal{C}(i:j)$ is the contangle of subsystem $i$ and $j$, defined by squared logarithmic negativity. We compare the deference between the contangle and $\xi$ measure through Fig. 3c and Fig. 6. It turns out the contangle measure satisfies the monogamy inequality but fails to show the conservation of entanglement. Thus it is better to apply $\xi$ to redefine the residual entanglement as $\mathcal{R}(A:BC)=\xi(A:BC)-\xi(A:B)-\xi(A:C)$. This development would help to the research on tripartite entanglement, which is our future interests.
\par
\begin{figure}
\begin{center}
\includegraphics[width=7cm,height=5.5cm]{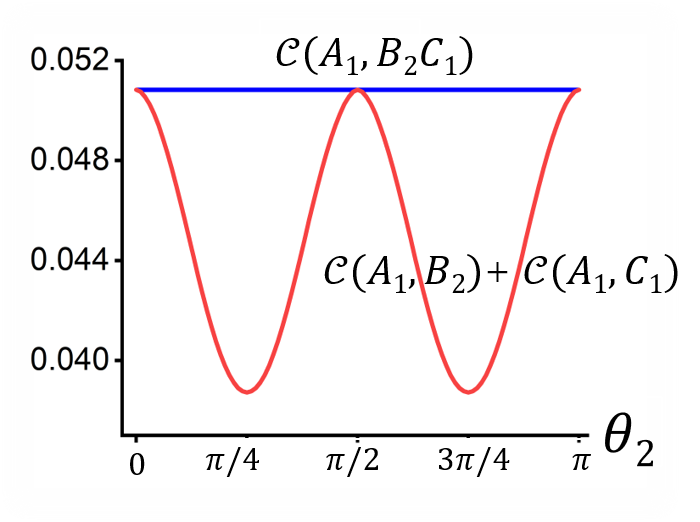}
\caption{
$\mathcal{C}(A_1,B_2C_1)$ and the summation of $\mathcal{C}(A_1,B_2)+\mathcal{C}(A_1,C_1)$ versus $\theta_2$. All the parameters are employed the same as in Fig. 3, which are listed in Sec.\uppercase\expandafter{\romannumeral3} C.}
\end{center}
\end{figure}
\indent Moreover, with Eq. (\ref{eq:xi-def}), it implies $\xi$ can be used beyond CV systems, for example, in determining the entanglement of discontinuous q-bits. $\xi$ has the potential to be a general entanglement measure which may deserve further studies on its properties, such as monotonicity under local operations and classical communications.   

\section*{Acknowledgements}
We would like to thank Yusef Maleki for helpful discussions. This research is supported by the project NPRP 13S-0205-200258 of the Qatar National Research Fund (QNRF) One of us (Liu J) is grateful to HEEP for financial support.
\section*{\centering Appendix A: Covariance matrix of Gaussian state under Bosonic field variables}
Under quadrature field variables $(q,p)$, the characteristic function of Gaussian state is given by \cite{rendell2005entanglement,wang2007quantum}
\begin{equation*}
    \chi(\boldsymbol{x})=\exp[-\frac12\boldsymbol{x}^T\gamma \boldsymbol{x}-id^T\boldsymbol{x}], \tag{A1}
\end{equation*}
where $\boldsymbol{x}^T=(q,p)$ and with the definition of anti-commutator $\{\hat{X},\hat{P}\}=\hat{X}\hat{P}+\hat{P}\hat{X}$,
\begin{equation*} \label{A2}
    \gamma =\left[
    \begin{matrix}
    \langle\hat{X}^2\rangle & \langle\frac12\{\hat{X},\hat{P}\}\rangle \\
    \langle\frac12\{\hat{X},\hat{P}\}\rangle & \langle\hat{P}^2\rangle
    \end{matrix}
    \right],
    \quad d=(\langle\hat{X}\rangle,\langle\hat{P}\rangle). \tag{A2}
\end{equation*}
The vector $d^T$ does not play a significant role in determining the entanglement since it only describes the average value of space and momentum, so we set it as zero without loss of generality.
\par
We transfer space-momentum operators to Bosonic field operators with the relation
\begin{equation*}
    \hat a=1/\sqrt{2}(\hat x + i \hat P), \quad \hat a^\dag=1/\sqrt{2}(\hat x - i \hat P). \tag{A3}
\end{equation*}
It then follows
\begin{equation*}
    \left[
    \begin{matrix}
    q\\
    p
    \end{matrix}
    \right]
    =\frac1{\sqrt{2}}
    \left[
    \begin{matrix}
    1&1\\
    -i&i
    \end{matrix}
    \right]
    \left[
    \begin{matrix}
    \alpha\\
    \alpha^*
    \end{matrix}
    \right]. \tag{A4}
\end{equation*}
The characteristic function is rewritten by
\begin{equation*}
\begin{aligned}
    &\chi(\alpha^*,\alpha)=\chi(q,p)=\exp(-\frac12\boldsymbol{x}^T\gamma \boldsymbol{x})\\
    =&\exp(-\frac12[\alpha^*,\alpha]\frac12\left[
    \begin{matrix}
    1&i\\
    1&-i
    \end{matrix}
    \right]\gamma\left[
    \begin{matrix}
    1&1\\
    -i&i
    \end{matrix}
    \right]\left[
    \begin{matrix}
    \alpha\\
    \alpha^*
    \end{matrix}
    \right]) \\
    =&\exp(-\frac12\boldsymbol{x'}^\dag V \boldsymbol{x'}).
\end{aligned} \tag{A5}
\end{equation*}
Therefore, instead of $\boldsymbol{x}$, under Bosonic field variables vector $\boldsymbol{x'}=(\alpha^*,\alpha)^\dag$, the covariance matrix $V$ is related to $\gamma$ by
\begin{equation*}
\begin{aligned}
    V&=\frac12\left[
    \begin{matrix}
    1&i\\
    1&-i
    \end{matrix}
    \right]\gamma\left[
    \begin{matrix}
    1&1\\
    -i&i
    \end{matrix}
    \right]\\
    &=\frac12\left[
    \begin{matrix}
    \gamma_{11}+\gamma_{22}&\gamma_{11}-\gamma_{22}+2i\gamma_{12}\\
    \gamma_{11}-\gamma_{22}-2i\gamma_{12}&\gamma_{11}+\gamma_{22}
    \end{matrix}
    \right]\\
    &=\left[
    \begin{matrix}
    a&b\\
    b^*&a
    \end{matrix}
    \right],
\end{aligned} \tag{A6}
\end{equation*}
where $\gamma_{ij}$ has been presented in Eq. (\ref{A2}).
\section*{\centering Appendix B: Important theorems}
In this paper, all the two-dimensional matrices during the calculation are symmetric matrix, and their diagonal values are equal to each other. Some tricks when manipulating these matrices are frequently used throughout the following text.
\par
We consider two matrices $A$ and $B$ given by
\begin{equation*}
    A=\left[
    \begin{matrix}
    a&b\\
    b&a
    \end{matrix}
    \right],\ 
    B=\left[
    \begin{matrix}
    c&d\\
    d&c
    \end{matrix}
    \right].
\end{equation*}
(a). Theorem 1. It can be proved that
\begin{equation} \label{B1}
    AB=BA. \tag{B1}
\end{equation}
\\
(b). Theorem 2. Given that
\begin{equation*}
\begin{aligned}
|A+B|=&(a+c)^2-(b+d)^2\\
=&a^2-b^2+c^2-d^2+2(ac-bd)
\end{aligned}
\end{equation*}
where $|\cdot|$ stands for the determinant of a matrix. Since $|B|B^{-1}=\left[
    \begin{matrix}
    c&-d\\
    -d&c
    \end{matrix}
    \right]$, 
it follows
\begin{equation}
\begin{aligned}
    |A+B|&=|A|+|B|+\text{Tr}(AB^{-1})|B|\\
         &=|A|+|B|+\text{Tr}(BA^{-1})|A|.
\end{aligned}\tag{B2}
\end{equation}
\\
(c). Theorem 3. It can be proved that
\begin{equation}
\begin{aligned}
    |A+B|(A+B)^{-1}&=\left[
    \begin{matrix}
    a+c&-b-d\\
    -b-d&a+c
    \end{matrix}
    \right]\\
    &=\left[
    \begin{matrix}
    a&-b\\
    -b&a
    \end{matrix}
    \right]
    +\left[
    \begin{matrix}
    c&-d\\
    -d&c
    \end{matrix}
    \right]\\
    &=|A|A^{-1}+|B|B^{-1}.
\end{aligned}\tag{B3}
\end{equation}
\\
(d) Theorem 4.
$A,B,C$ are two-dimensional matrices and $A$ is reversible. Then
\begin{equation} \label{B4}
\begin{aligned}
    \text{Det}\left[
    \begin{matrix}
    A&B\\
    B&C
    \end{matrix}
    \right]=&|A|\cdot|C-BA^{-1}B| \\
    =&|AC-ABA^{-1}B|\\
    =&|AC-BB|.
\end{aligned}\tag{B4}.
\end{equation}
where the last step is based on Eq. (\ref{B1}).
\\
\\
(e) Theorem 5.
Define $A_1=\cos^2\theta\cdot A+\sin^2\theta\cdot B$, $B_1=\cos^2\theta\cdot B+\sin^2\theta\cdot A$, and $(AB)_1=(A-B)\sin\theta\cos\theta$, then
\begin{equation}
    A_1B_1-(AB)_1(AB)_1=AB. \tag{B5}
\end{equation}
Note that $(AB)_1$ is a matrix related to $A$ and $B$. As shown later in Appendix C, it denotes the off-diagonal matrix of $V_{\text{out}}$, which is different from $AB$, the matrix product of $A$ and $B$.
\section*{\centering Appendix C: Entanglement and residual nonclassicality of two-mode Gaussian state}
Substituting Eq. (\ref{U}) into Eq. (\ref{Vout}), 
\begin{equation}
    V_\text{out}=\left[
    \begin{matrix}
    A_1&(AB)_1\\
    (AB)^\dag_1&B_1
    \end{matrix}
    \right], \tag{C1}
\end{equation}
where the matrix $A_1,B_1,(AB)_1$ are given by
\begin{equation}
\begin{aligned}
&A_1=\left[
    \begin{matrix}
    a\cos^2\theta+c\sin^2\theta & b\cos^2\theta+d\sin^2\theta e^{2i\varphi}\\
    b^*\cos^2\theta+d^*\sin^2\theta e^{-2i\varphi} &a\cos^2\theta+c\sin^2\theta
    \end{matrix}
    \right],\\
    \\
&B_1=\left[
    \begin{matrix}
    a\sin^2\theta+c\cos^2\theta&b\sin^2\theta e^{-2i\varphi}+d\cos^2\theta \\
    b^*\sin^2\theta e^{2i\varphi}+d^*\cos^2\theta &a\sin^2\theta+c\cos^2\theta
    \end{matrix}
    \right],\\
    \\
&(AB)_1=\left[
    \begin{matrix}
    (a-c)e^{i\varphi} & be^{-i\varphi}-de^{i\varphi}\\
    b^*e^{i\varphi}-d^*e^{-i\varphi} &(a-c)e^{-i\varphi}
    \end{matrix}
    \right]\sin\theta\cos\theta.
\end{aligned} \tag{C2}
\end{equation}
\\
The constraint in Eq. (\ref{constraints_1}b), which is $\varphi=\frac12[\text{arg}(b)-\text{arg}(d)]$, make it possible to simplify $b\cos^2\theta+d\sin^2\theta e^{2i\varphi}$. It leads us to redefine $b=|b|, d=|d|$, and $\varphi=0$. We have verified that without loss of generality, all the results will remain unchanged in the following derivation. It then follows
\begin{equation} \label{C3}
\begin{aligned}
A_1&=A\cos^2\theta+B\sin^2\theta,\\
B_1&=A\sin^2\theta+B\cos^2\theta,\\
(AB)_1&=(A-B)\sin\theta\cos\theta.
\end{aligned} \tag{C3}
\end{equation}
As we mentioned in Sec.\uppercase\expandafter{\romannumeral2} A, $\lambda$ denotes the minimum eigenvalue of a matrix. With the above expression of $A_1,B_1$, we obtain the transformation relation of $\lambda_{A_1}$, $\lambda_{B_1}$ and $\lambda_{A}$, $\lambda_{B}$ as
\begin{equation} \label{C4}
\begin{aligned}
    \lambda_{A_1}&=\cos^2\theta\cdot\lambda_A + \sin^2\theta\cdot\lambda_B, \\
    \lambda_{B_1}&=\sin^2\theta\cdot\lambda_A + \cos^2\theta\cdot\lambda_B.
\end{aligned} \tag{C4}
\end{equation}
After several transformation, Eq. (\ref{C4}) leads to
\begin{equation*} \label{C5}
\frac{\lambda_{A_1}\lambda_{B_1}}{\lambda_A\lambda_B}-1=\frac{(\lambda_B-\lambda_A)^2}{4\lambda_A\lambda_B}\sin^2(2\theta). \tag{C5}  
\end{equation*} 
Considering $S_\mathcal{N}=\log_2(\lambda_{A_1}\lambda_{B_1})/(\lambda_A\lambda_B)$, 
\begin{equation*}
    S_\mathcal{N}>0\quad\text{is equivalent to}\quad \frac{\lambda_{A_1}\lambda_{B_1}}{\lambda_A\lambda_B}-1>0. \tag{C6}
\end{equation*}
On the other hand, as mentioned in Eq. (\ref{xi_expression}), $E_\mathcal{N}>0$ is equivalent to $\xi>0$. The new quantifier $\xi=S-1/2-8\text{Det}[V_\text{out}]$ can be described in terms of $\lambda_A, \lambda_B$ as well.
\begin{equation}
\begin{aligned}
S&=2\ \text{Det}[A_1]+2\ \text{Det}[B_1]-4\ \text{Det}[(AB)_1]\\
&=2(\lambda_{A_1}\Lambda_{A_1}+\lambda_{B_1}\Lambda_{B_1})-[(a-c)^2-(b-d)^2]\sin^2(2\theta) \\
&=2(\lambda_{A}\Lambda_{A}+\lambda_{B}\Lambda_{B})(1-2\sin^2\theta\cos^2\theta)+4(\lambda_{A}\Lambda_{B}
\\&+\lambda_{B}\Lambda_{A})\cdot\sin^2\theta\cos^2\theta-(\lambda_A-\lambda_B)(\Lambda_A-\Lambda_B)\sin^2(2\theta)\\
&=2(\lambda_{A}\Lambda_{A}+\lambda_{B}\Lambda_{B})-2(\lambda_A-\lambda_B)(\Lambda_A-\Lambda_B)\sin^2(2\theta),
\end{aligned} \tag{C7}
\end{equation}
\\
where $\Lambda$ is the maximum eigenvalue of a CM in contrast to the minimum eigenvalue $\lambda$. Recall that $\text{Det}[V_\text{out}]=|A|\cdot|B|=\lambda_A\Lambda_A\lambda_B\Lambda_B$ and both $A,B$ are pure states ($\lambda_A\Lambda_A=\lambda_B\Lambda_B=1/4$), it then follows 
\begin{equation}
\begin{aligned}
&S-\frac12-8\ \text{Det}[V_\text{out}]=S-\frac12-8\times\frac14\times\frac14\\
&=2(\lambda_B-\lambda_A)(\Lambda_A-\Lambda_B)\sin^2(2\theta)\\
&=\mathcal{C}\cdot(\frac{\lambda_{A_1}\lambda_{B_1}}{\lambda_A\lambda_B}-1).
\end{aligned} \tag{C8}
\end{equation} 
Equation (\ref{C5}) is applied in the last step where 
\begin{equation}
\mathcal{C}=8\ \lambda_A\lambda_B\frac{\Lambda_A-\Lambda_B}{\lambda_B-\lambda_A}. \tag{C9}
\end{equation}
Recall that the expression for $\lambda_A,\Lambda_A,\lambda_B,\Lambda_B$ are given by
\begin{equation*}
\begin{aligned}
\lambda_A&=a-|b|=\sqrt{1/4+|b|^2}-|b|, \\
\Lambda_A&=a+|b|=\sqrt{1/4+|b|^2}+|b|, \\
\lambda_B&=c-|d|=\sqrt{1/4+|d|^2}-|d|, \\
\Lambda_B&=c+|d|=\sqrt{1/4+|d|^2}+|d|. \\
\end{aligned} \tag{C10}
\end{equation*}
Note that $\sqrt{1/4+x^2}-x$ is a monotonically decreasing function while $\sqrt{1/4+x^2}+x$ is a monotonically increasing one, which leads to $\lambda_A<\lambda_B\leqslant\Lambda_B<\Lambda_A$ or $\lambda_B<\lambda_A\leqslant\Lambda_A<\Lambda_B$. In both cases, $\mathcal{C}>0$.
\par
As can be seen from the above expression, $S_\mathcal{N}>0$ is a necessary and sufficient condition for $S-\frac12-8\ \text{Det}[V_\text{out}]>0$, which means the two-mode Gaussian entanglement exists.

\section*{\centering Appendix D: Distribution relation of entanglement in a two-BS system}
As shown by Fig. 2, three Gaussian states are mixed by a two-BS network. Their output CM is given by
\begin{equation}
    V_\text{out}=U^\dag(\theta_2,\varphi_2)U^\dag(\theta_1,\varphi_1)V_\text{in}U(\theta_1,\varphi_1)U(\theta_2,\varphi_2). \tag{D1}
\end{equation}
For $\xi(A_1,B_2)$, we trace out mode $C$ on the output state in order to obtain the covariance matrix $V_{A_1B_2}$ of the state $\rho_{A_1B_1}$. It is given that
\begin{equation}
\begin{aligned}
V_{A_1B_2}=&\left[
\begin{matrix}
A_1 & (AB)_1\cos\theta_2\\
(AB)_1\cos\theta_2 & B_1\cos^2\theta+C\sin^2\theta_2
\end{matrix}
\right],\\
C=&\left[
\begin{matrix}
g & f\\
f & g
\end{matrix}
\right],
\end{aligned} \tag{D2}
\end{equation}
where $A_1, (AB)_1, B_1$ are given by Eq. (\ref{C3}). Following the idea introduced in Appendix C, we redefine $\varphi_2=0$ and $f=|f|$ when calculating the determination of matrix.
\\
\\
For $\xi(A_1,B_2)=S_{A_1B_2}-1/2-8|V_{A_1B_2}|$, 
\begin{equation} \label{D3}
    S_{A_1B_2}=2|A_1|+2|B_1\cos^2\theta_2+C\sin^2\theta_2|-4|(AB)_1|\cos^2\theta_2, \tag{D3}
\end{equation}
where the second term $|B_1\cos^2\theta_2+C\sin^2\theta_2|$ is calculated based on Theorem 4
\begin{equation} \label{D4}
\begin{aligned}
&|B_1\cos^2\theta_2+C\sin^2\theta_2|\\
=&|B_1|\cos^4\theta_2+|C|\sin^4\theta_2+\sin^2\theta_2\cos^2\theta_2\text{Tr}(B_1C^{-1})|C|.
\end{aligned} \tag{D4}
\end{equation}
For $|V_{A_1B_2}|$, according to Theorem 4 or Eq. (\ref{B4}), 
\begin{equation}
\begin{aligned}
    &|V_{A_1B_2}|\\
    =&|A_1(B_1\cos^2\theta+C\sin^2\theta_2)-(AB)_1(AB)_1\cos^2\theta_2|\\
    =&|\cos^2\theta_2[A_1B_1-(AB)_1(AB)_1]+\sin^2\theta_2A_1C|\\
    =&\cos^4\theta_2|V_{A_1B_1}|+\sin^4\theta_2|A_1||C|+\\
    &\cos^2\theta_2\sin^2\theta_2\text{Tr}([A_1B_1-(AB)_1(AB)_1]C^{-1}A_1^{-1})|A_1||C|\\
    =&\cos^4\theta_2|V_{A_1B_1}|+\frac14\sin^4\theta_2|A_1|+\\
    &\cos^2\theta_2\sin^2\theta_2\text{Tr}(AB|A_1|A_1^{-1}C^{-1})|C|.
\end{aligned} \tag{D5}
\end{equation}
Theorems $2,5$ are applied in the third and last step, respectively. According to Theorem 3,
\begin{equation}
\begin{aligned}
    AB|A_1|A^{-1}_1&=AB(|A|A^{-1}\cos^2\theta_1+|B|B^{-1}\sin^2\theta_1)\\
    &=|A|B\cos^2\theta_1+|B|A\sin^2\theta_1\\
    &=\frac14B.
\end{aligned} \tag{D6}
\end{equation}
Note that we apply the condition that $|A|=|B|=\frac14$ in the last step. Substituting it into the expression of $|V_{A_1B_2}|$,
\begin{equation} \label{D7}
\begin{aligned}
    |V_{A_1B_2}|=\cos^4\theta_2|V_{A_1B_1}|+\frac14\sin^4\theta_2|A_1|+\\
    \qquad\frac14\cos^2\theta_2\sin^2\theta_2\text{Tr}(BC^{-1})|C|.
\end{aligned} \tag{D7}
\end{equation}
As can be seen, $S_{A_1B_2}$ and $|V_{A_1B_2}|$ are split into many terms which are related to $S_{A_1B_1}$ and $|V_{A_1B_1}|$.  Substituting Eq. (\ref{D3}, \ref{D4}, \ref{D7}) to the expression of $\xi(A_1,B_2)$,
\begin{equation} \label{D8}
\begin{aligned}
    &\xi(A_1,B_2)=S_{A_1B_2}-1/2-8|V_{A_1B_2}|\\
    =&\cos^4\theta_2\cdot(S_{A_1B_1}-1/2-8|V_{A_1B_1}|)\ +\\
    &\qquad\qquad \sin^2\theta_2\cos^2\theta_2\cdot(4|A_1|-1-4|(AB)_1|).
\end{aligned} \tag{D8}
\end{equation}
Since $|A|=|B|=1/4$, we calculate $|A_1|$ as
\begin{equation}
\begin{aligned}
|A_1|&=|A\cos^2\theta_1+B\sin^2\theta_1|\\
&=|A|\cos^4\theta_1+|B|\sin^4\theta_1+\sin^2\theta_1\cos^2\theta\text{Tr}(AB^{-1})|B|\\
&=|A|\sin^4\theta_1+|B|\cos^4\theta_1+\sin^2\theta_1\cos^2\theta\text{Tr}(AB^{-1})|B|\\
&=|A\sin^2\theta_1+B\cos^2\theta_1|=|B_1|.
\end{aligned} \tag{D9}
\end{equation}
Recall that $|V_{A_1B_1}|=|V_{AB}|=|A||B|=\frac14\times\frac14$, it follows
\begin{equation} \label{D10}
\begin{aligned}
&4|A_1|-1-4|(AB)_1|\\
=&2|A_1|+2|B_1|-4|(AB)_1|-1/2-8\times|V_{A_1B_1}|\\
=&S_{A_1B_1}-1/2-8|V_{A_1B_1}|. 
\end{aligned} \tag{D10}
\end{equation}
Substituting Eq. (\ref{D10}) into Eq. (\ref{D8}),
\begin{equation} \label{D11}
\begin{aligned}
&S_{A_1B_2}-1/2-8|V_{A_1B_2}|\\
    =&(\cos^4\theta_2+\sin^2\theta_2\cos^2\theta_2)\cdot(S_{A_1B_1}-1/2-8|V_{A_1B_1}|)\\
    =&\cos^2\theta_2\cdot(S_{A_1B_1}-1/2-8|V_{A_1B_1}|)\\
    =&\cos^2\theta_2\cdot\xi(A_1,B_1).
\end{aligned} \tag{D11}
\end{equation}
With the above calculation, we obtain
\begin{equation}
\xi(A_1,B_2)=\cos^2\theta_2\cdot \xi(A_1,B_1). \tag{D12}
\end{equation}
Using the same steps to calculate $\xi(A_1,C_1)$, it can be obtained that
\begin{equation}
\xi(A_1,C_1)=\sin^2\theta_2\cdot \xi(A_1,B_1), \tag{D13}
\end{equation}
which means entanglement could be distributed as the same way of distributing reflectance and transmittance. 
\section*{\centering Appendix E: Trace norm equality relations}
$\rho_{A_1:B_1}^{T_A}$ can be written as $\rho_{A_1:B_1}^{T_A}=\sum_{i}\lambda_i\rho_i$ \cite{vidal2002computable}, where $\rho_i=\ket{\psi_{AB}}_i\bra{\psi_{AB}}_i$ stands for a series of pure states. $\rho_i\rho_j=\delta_{ij}\rho_i$.
\begin{equation*}
\begin{aligned}
    \sqrt{(\rho_{A_1:B_1}^{T_A})(\rho_{A_1:B_1}^{T_A})\dagger}&=\sqrt{\sum_{i,j}\lambda_i\lambda_j^*\rho_i\rho_j}=\sqrt{\sum_{i}|\lambda_i|^2\rho_i}\\
    &=\sum_i|\lambda_i|\rho_i
\end{aligned} \tag{E1}
\end{equation*}
It leads to $||\rho_{A_1:B_1}^{T_A}||_1=\text{Tr}\left[\sqrt{(\rho_{A_1:B_1}^{T_A})(\rho_{A_1:B_1}^{T_A})\dagger}\right]=\sum_i|\lambda_i|$.
According to Fig. 2, $\rho_{A_1:B_2C_1}$ is obtained by a unitary transformation on $\rho_{A_1:B_1}\otimes\rho_C$, recall that $\rho_c$ is a pure state,
\begin{equation*}
    \rho_{A_1:B_2C_1}=u_{BC}^\dagger(\rho_{A_1:B_1}\otimes\rho_C)u_{BC}. \tag{E2}
\end{equation*}
Thus,
\begin{equation*}
\begin{aligned}
    \rho_{A_1:B_2C_1}^{T_A}&=u_{BC}^\dagger(\rho_{A_1:B_1}^{T_A}\otimes\rho_C)u_{BC}.\\
    &=u_{BC}^\dagger(\sum_i\lambda_i\rho_i\otimes\rho_C)u_{BC}.
\end{aligned} \tag{E3}
\end{equation*}
It follows that
\begin{equation*}
\begin{aligned}
    \sqrt{(\rho_{A_1:B_2C_1}^{T_A})(\rho_{A_1:B_2C_1}^{T_A})\dagger}&=\sqrt{u_{BC}^\dagger(\sum_{i,j}\lambda_i\lambda_j^*\rho_i\rho_j\otimes\rho_C)u_{BC}}\\
    &=\sqrt{u_{BC}^\dagger(\sum_{i}|\lambda_i|^2\rho_i\otimes\rho_C)u_{BC}}\\
    &=u_{BC}^\dagger(\sum_{i}|\lambda_i|\cdot\rho_i\otimes\rho_C)u_{BC}.
\end{aligned} \tag{E4}
\end{equation*}
The trace norm of $\rho_{A_1:B_2C_1}^{T_A}$ is calculated by
\begin{equation*}
    ||\rho_{A_1:B_2C_1}^{T_A}||_1=\text{Tr}\left[u_{BC}^\dagger(\sum_{i}|\lambda_i|\cdot\rho_i\otimes\rho_C)u_{BC}\right]=\sum_i|\lambda_i|, \tag{E5}
\end{equation*}
which means $||\rho_{A_1:B_2C_1}^{T_A}||_1=||\rho_{A_1:B1}^{T_A}||_1$. 
\par
The second equation, $||\rho_{A_1:B_2C_1}^{2}||_1=||\rho_{A_1:B_1}^{2}||_1$ can be obtained in the similar method.
\section*{\centering Appendix F: Distribution relation of entanglement in a linearly arranged BS system}
A general proof of the distribution relation in a linearly arranged $n$-BS system is presented in this section (Fig. 5).
\par
Define
\begin{equation}
V_1=\left[
\begin{matrix}
A_1 & (AB)_1\\
(AB)_1 & B_1
\end{matrix}
\right],\quad \xi_1=S_{A_1B_1}-1/2-8|V_1|,   \tag{F1} 
\end{equation}
where $V_1$ is the covariance matrix of $\rho_{A_1B_1}$. The one for  $\rho_{A_1B_2}$, $V_2$, is related to $V_1$ as 
\begin{equation}
\begin{aligned}
V_2&=\left[
\begin{matrix}
A_1 & (AB)_2\\
(AB)_2 & B_2
\end{matrix}
\right],   \\
(AB)_2&=(AB)_1\cos\theta_2, \\
B_2&=B_1\cos^2\theta_2+\frac12\sin^2\theta_2,
\end{aligned} \tag{F2} 
\end{equation}
where $\frac12$ (to be specific, $\frac12\mathbf{I_2}$) stands for a new input state, which is vacuum or coherent states as we mentioned before. The same relationship works for $V_n$ and $V_{n-1}$ as
\begin{equation}
\begin{aligned}
V_n&=\left[
\begin{matrix}
A_1 & (AB)_n\\
(AB)_n & B_n
\end{matrix}  
\right], \\
(AB)_n&=(AB)_{n-1}\cos\theta_n, \\
B_n&=B_{n-1}\cos^2\theta_n+\frac12\sin^2\theta_n.
\end{aligned} \tag{F3}
\end{equation}
Following the same way of Eq. (\ref{D3}-\ref{D11}),
\begin{equation} \label{F4}
\begin{aligned}
    S_{A_1B_n}=2|A_1|-4|(AB)_{n-1}|\cos^2\theta_n+2|B_{n-1}|\cos^4\theta_n\\
    +2\times\frac14\sin^4\theta_n+2\times\frac12\sin^2\theta_n\cos^2\theta_n\text{Tr}(B_{n-1}).
\end{aligned} \tag{F4} 
\end{equation}
On the other hand,
\begin{equation} \label{F5}
\begin{aligned}
    &|V_{A_1B_n}|=\cos^4\theta_n|V_{A_1B_{n-1}}|+\frac14\sin^4\theta_n|A_1|+\\
    &\frac12\sin^2\theta_n\cos^2\theta_n\text{Tr}(|A_1|A_1^{-1}[A_1B_{n-1}-(AB)_{n-1}(AB)_{n-1}]).
\end{aligned} \tag{F5} 
\end{equation}
Substituting Eq. (\ref{F4}, \ref{F5}) into the expression of $\xi_n$,
\begin{equation} \label{F6}
\begin{aligned}
&\xi_n=S_{A_1B_n}-1/2-8|V_{A_1B_n}|\\
&=\cos^4\theta_n\xi_{n-1}+\sin^2\theta_n\cos^2\theta_n\left(4|A_1|-1-4|(AB)_{n-1}|\right)+\\
&\sin^2\theta_n\cos^2\theta_n\text{Tr}(B_{n-1}-4|A_1|A_1^{-1}[A_1B_{n-1}-(AB)_{n-1}(AB)_{n-1}]).
\end{aligned} \tag{F6} 
\end{equation}
The complex term in the trace operator bracket, $B_{n-1}-4|A_1|A_1^{-1}[A_1B_{n-1}-(AB)^2_{n-1}]$, can be derived mathematically by the idea of iteration 
\begin{equation}  
\begin{aligned}
n&=2,\ B_{1}-4|A_1|A_1^{-1}[A_1B_{1}-(AB)^2_{1}]\\
&=B_1-4|A_1|A_1^{-1}[A_1B_1-(AB)_1(AB)_1])\\
&=B_1-4\times\frac14B_1=0,\\
n&=3,\ B_{2}-4|A_1|A_1^{-1}[A_1B_{2}-(AB)^2_{2}]\\
&=\cos^2\theta_2\cdot0+\sin^2\theta_2(\frac12-2|A_1|)\cdot\mathbf{I_2}\\
&=(1-\cos^2\theta_2)(\frac12-2|A_1|)\cdot\mathbf{I_2},\\
n&=4,\ B_{3}-4|A_1|A_1^{-1}[A_1B_{3}-(AB)^2_{3}]\\
&=\cos^2\theta_3\cdot(1-\cos^2\theta_2)(\frac12-2|A_1|)\cdot\mathbf{I_2}+\sin^2\theta_3(\frac12-2|A_1|)\cdot\mathbf{I_2}\\
&=(1-\cos^2\theta_2\cos^2\theta_3)\cdot(\frac12-2|A_1|)\cdot\mathbf{I_2}.
\end{aligned} \tag{F7} 
\end{equation}
It follows that
\begin{equation}
\begin{aligned}
&\text{Tr}\left[B_{n-1}-4|A_1|A_1^{-1}[A_1B_{n-1}-(AB)^2_{n-1}]\right]\\
=&\text{Tr}\left[(1-\prod_{i=2}^{n-1}\cos^2\theta_i)\cdot(\frac12-2|A_1|)\cdot\mathbf{I_2}\right]\\
=&(1-\prod_{i=2}^{n-1}\cos^2\theta_i)\cdot(1-4|A_1|).
\end{aligned} \tag{F8} 
\end{equation}
Substituting it into Eq. (\ref{F6})
\begin{equation}
\begin{aligned}
\xi_n=&\cos^4\theta_n\cdot\xi_{n-1}+\sin^2\theta_n\cos^2\theta_n \bigg\{ 4|A_1|-1-4|(AB)_{n-1}|+\\
&(1-\prod_{i=2}^{n-1}\cos^2\theta_i)(1-4|A_1|)\bigg\}\\
=&\cos^4\theta_n\cdot\xi_n+\sin^2\theta_n\cos^2\theta_n\bigg\{(\prod_{i=2}^{n-1}\cos^2\theta_i)(4|A_1|-1)-\\
&4|(AB)_{n-1}|\bigg\}.\\
\end{aligned} \tag{F9} 
\end{equation}
Recall that $(AB)_{n-1}=\cos\theta_{n-1}\cdot(AB)_{n-2}=\prod_{i=2}^{n-1}\cos\theta_i\cdot(AB)_1$, it follows that
\\
\begin{equation}
\begin{aligned}
\xi_n=&\cos^4\theta_n\cdot\xi_n+\sin^2\theta_n\cos^2\theta_n\bigg\{(\prod_{i=2}^{n-1}\cos^2\theta_i)(4|A_1|-1)-\\
&4(\prod_{i=2}^{n-1}\cos^2\theta_i)\cdot|(AB)_1|\bigg\}\\
=&\cos^4\theta_n\cdot\xi_{n-1}+\sin^2\theta_n(\prod_{i=2}^{n}\cos^2\theta_i)(4|A_1|-1-4|(AB)_1|)\\
=&\cos^4\theta_n\cdot\xi_{n-1}+\sin^2\theta_n\cdot(\prod_{i=2}^{n}\cos^2\theta_i)\cdot\xi_1.
\end{aligned} \tag{F10} 
\end{equation}
Applying the idea of iteration again,
\begin{equation} 
\begin{aligned}
n=2,\ \xi_2&=\cos^4\theta_2\cdot\xi_1+\sin^2\theta_2\cos^2\theta_2\cdot\xi_1=\cos^2\theta_2\cdot\xi_1,\\
n=3,\ \xi_3&=\cos^4\theta_3\cdot\xi_2+\sin^2\theta_3\cos^2\theta_3\cdot\cos^2\theta_2\cdot\xi_1\\
&=\cos^4\theta_3\cdot\xi_2+\sin^2\theta_3\cos^2\theta_3\cdot\xi_2=\cos^2\theta_3\cdot\xi_2.
\end{aligned} \tag{F11} 
\end{equation}
It follows that
\begin{equation}
\xi_n=\cos^2\theta_n\cdot\xi_{n-1}=\prod_{i=2}^n\cos^2\theta_i\cdot \xi_1. \tag{F12} 
\end{equation}
Thus, the distribution relation of entanglement in a linearly arranged $n$-BS system is proved.

\bibliography{ref}
\end{document}